\documentclass[12pt]{iopart}
\usepackage{graphicx}

\begin{document}

\title[Spurious Solar-Wind Effects on Acceleration Noise in LISA Pathfinder]{Spurious Solar-Wind Effects on Acceleration Noise in LISA Pathfinder}

\author{Arnold Yang\(^1\), Indie Desiderio-Sloane\(^1\)\(^2\), and Grant David Meadors\(^3\)}

\address{\(^1\)Institute for Computing in Research, 145 Washington Ave, Santa Fe, NM 87501}
\address{\(^2\)California Institute of Technology, 1200 East California Boulevard Pasadena, California 91125, United States}
\address{\(^3\)Los Alamos National Laboratory, Los Alamos National Laboratory P.O. Box 1663 Los Alamos, NM 87545}
\ead{gdmeadors@lanl.gov}
\vspace{10pt}
\begin{indented}
\item[]\today
\end{indented}

\begin{abstract}
Spurious solar-wind effects are a potential noise source in the measurements of the future Laser Interferometer Space Antenna (LISA). Comparative models are used to predict the possible impact of this noise factor and estimate spurious solar-wind effects on acceleration noise in LISA Pathfinder (LPF). Data from NASA’s Advanced Composition Explorer (ACE), situated at the L1 Lagrange point, served as a reliable source of solar-wind data. The data sets were compared over the 114-day time period from March 1, 2016 to June 23, 2016. To evaluate these effects, the data from both satellites were formatted, gap-filled, and adapted for comparison, and a coherence plot comparing the results of the Fast Fourier Transformations. The coherence plot suggested that solar-wind had a minuscule effect on the LPF, and higher frequency coherence (LISA’s main observing band) can be attributed to random chance correlation. This result indicates that measurable correlation due to solar-wind noise over 3-month timescales can be ruled out as a noise source. This is encouraging, although another source of noise from the sun, solar irradiance pressure, is estimated to have a more significant effect and has yet to be analyzed.
\end{abstract}

%
\vspace{2pc}
\noindent{\it Keywords}: Space weather, solar wind, gravitational waves, LISA
%
%
%
%

\section{Introduction}
The future Laser Interferometer Space Antenna (LISA) will be the first space-based gravitational wave detector on a heliocentric orbit \cite{amaro2017laser}. The LISA project expands upon the work of the ground-based Laser Interferometer Gravitational-wave Observatory (LIGO), which has made significant contributions to our understanding of the universe \cite{LIGO2015}. LISA will measure frequencies ranging from approximately 0.1 millihertz to 100 millihertz \cite{amaro2017laser}.  LISA will consist of three spacecraft in an equilateral triangle with arms 2.5 million km long. The antenna will follow 20 degrees behind the Earth. Each spacecraft will contain two centered, free-falling test masses. Deviations from the geodesic motion of the test masses are detected with time-delay interferometry (TDI) \cite{amaro2017laser}, using six laser links that connect each of the three spacecraft bidirectionally. TDI constructs a set of time series data, sampled every few seconds, from the combination of these laser links that represents that stretching of space, known as (dimensionless) strain. LISA's TDI output is, in effect, three virtual, correlated space-based interferometer channels sensitive to gravitational waves in the millihertz band.

\begin{figure}[!ht]
\centerline{\includegraphics[width=0.9\textwidth]{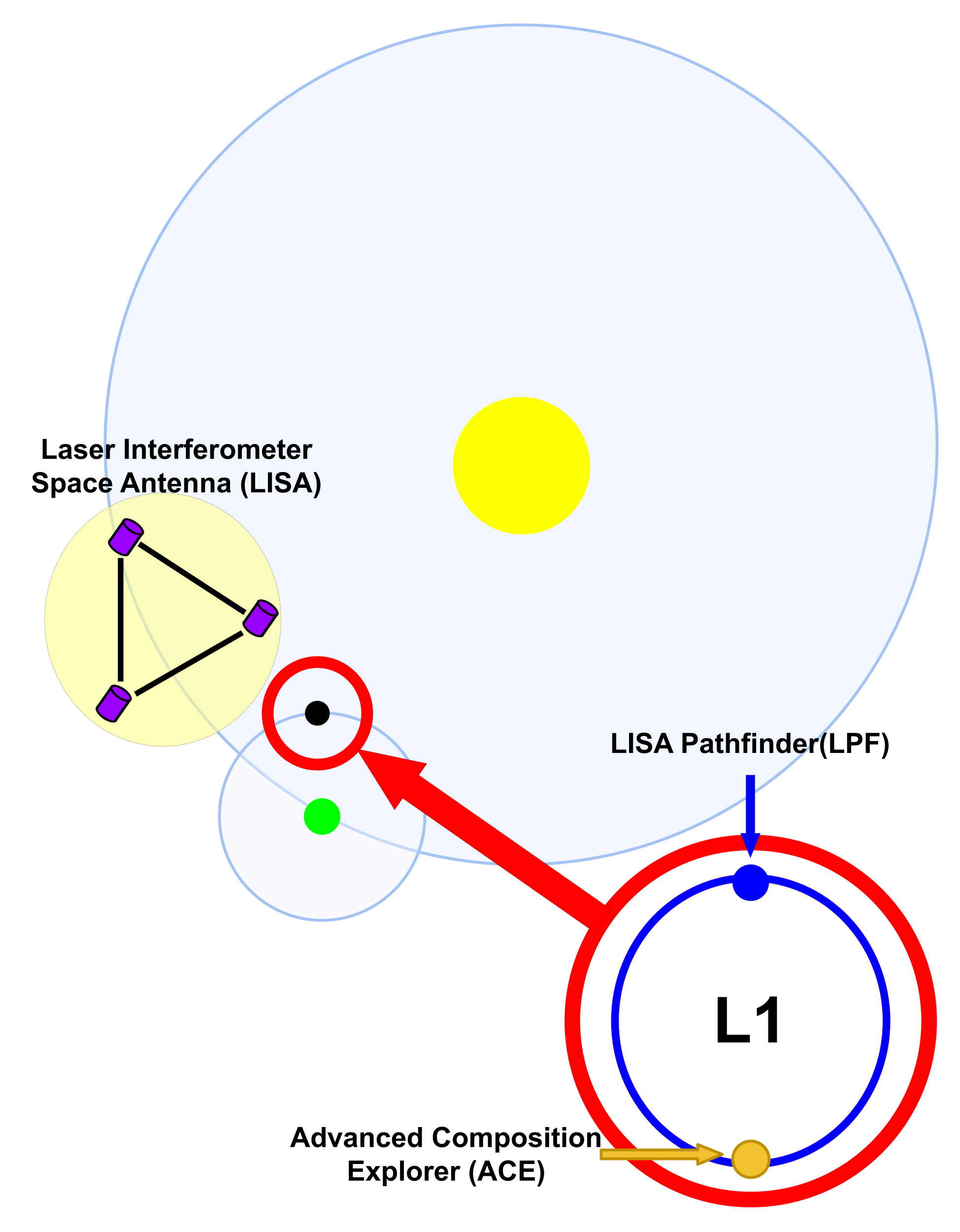}}
	\caption{This diagram depicts the positioning of the LISA Pathfinder in relation to the Sun and Earth. The planned LISA system will be 20\textdegree{} behind the Earth.}
\label{Figure 1}
\end{figure}

To confirm the viability of the LISA concept, the European Space Agency (ESA) developed and launched the LISA Pathfinder (LPF). LPF consists of a test version of a LISA module - two free acceleration test masses held in the module's center. The LPF was launched on December 3, 2015, and traveled 1.5 million km from Earth to orbit the L1 Lagrange point; see Figure 1 for positioning in relation to the Sun and Earth. The goal of LPF is to show that a drag-free control measurement technique is viable for a long baseline gravitational wave observatory. LPF was sent into space with 2 test masses, the second one located 38 cm from the first, to avoid common noise sources: solar radiation pressure, solar-wind, magnetic environment, and particle impacts. In addition, spacecraft acceleration was measured to observe external disturbances on a single-test mass signal. One of these sources, solar-wind, has been generally understood to be a very modest noise source \cite{shaul2006solar}, below other factors - actuation fluctuations, Brownian noise, micrometeoroids, and more.

Theoretical extrapolation from solar wind and solar irradiance data to LISA \cite{frank2020modeling} has set the stage for the analysis in this manuscript, which considers whether correlations are measurable between solar wind and LPF acceleration data. Analyzing LPF's data can be proof that solar-wind noise impact is minimal or demonstrate that in the future, LISA might act as a serendipitous solar-wind sensor at an unusual orbit, similar to STEREO \cite{kaieser-et-al-2008}. Constant solar wind, sunlight radiation pressure, or other static forces are not expected to have an impact on LISA. However, a time-varying force could accelerate the LISA spacecraft, particularly if the three spacecraft experience the differential forces due to orbital position and orientation. Studies of solar wind variability~\cite{arge1999} demonstrate that the solar wind can vary from 350 km/s to 700 km/s in the span of a week, doubling the solar-wind force on a timescale of less than 1 Mega-second, motivating an investigation of the micro-Hertz to milli-Hertz frequencies even though they are below the typical LISA science band.

The idea was tested by comparing the LPF spacecraft's Z and X axis data with the corresponding ACE solar-wind data from March 1, 2016, to June 23, 2016 \cite{aceswepamlevel2data}. The methods used can find direct, linear couplings at the measured timescales, but nonlinear ``upconversion" of solar-wind noise to higher frequencies cannot be ruled out. The Advanced Composition Explorer (ACE), orbiting the L1 point as the LPF did, is used to measure solar-wind data, such as proton speed, proton density, proton to alpha particle ratio, $x, y, z$ components of proton velocity, and Coordinated Universal Time (UTC) \cite{aceswepamlevel2data}. This data can be modeled onto LPF and compared with the data from LPF to analyze the effects of solar-wind.

\subsection{Related Works}
Space weather has been identified as a key noise factor in possible future LISA measurements. Past work that assessed the impact of solar wind involved using data from the Solar and Heliospheric Observatory’s (SOHO) Virgo Solar irradiance and the ACE Solar-wind to model possible impacts on the future LISA mission \cite{frank2020modeling}. Another major source of noise identified for LPF is micrometeoroids. Through various models, 54 impact candidates were identified across a 4348-hour time period. They were shown to be similar to those resulting from Jupiter Family Comets (JFC), Oort cloud comets, Halley-type comets, and asteroids \cite{thorpe2019micrometeoroid}. Additionally, the development of LISA is renewing the interest in the environment of space \cite{Grimani_2006}\cite{10.1093/mnras/stac316}\cite{smetana2020background}. This environment includes solar energetic particles, galactic cosmic ray fluxes, and the effects of solar neutrons and interplanetary electrons \cite{shaul2006solar}.

\section{Methods}
 Figure 2 shows the steps to plot and compare the ACE and LPF data sets.
 We first formatted both data sets to include only data present is within the time frame and for the axes being studied. Next, we filter and gap-fill the ACE data set because the data set is in time-series format. An Inverse Fast Fourier Transform (IFFT) is applied to the LPF data set because the set is in the frequency domain \cite{harris2020array}. Once the IFFT is performed, the LPF data is filtered and gap-filled. The ACE data set must be modeled onto LPF. Lastly, both data sets are plotted and compared.
\begin{figure}
\centerline{\includegraphics[width=0.9\textwidth]{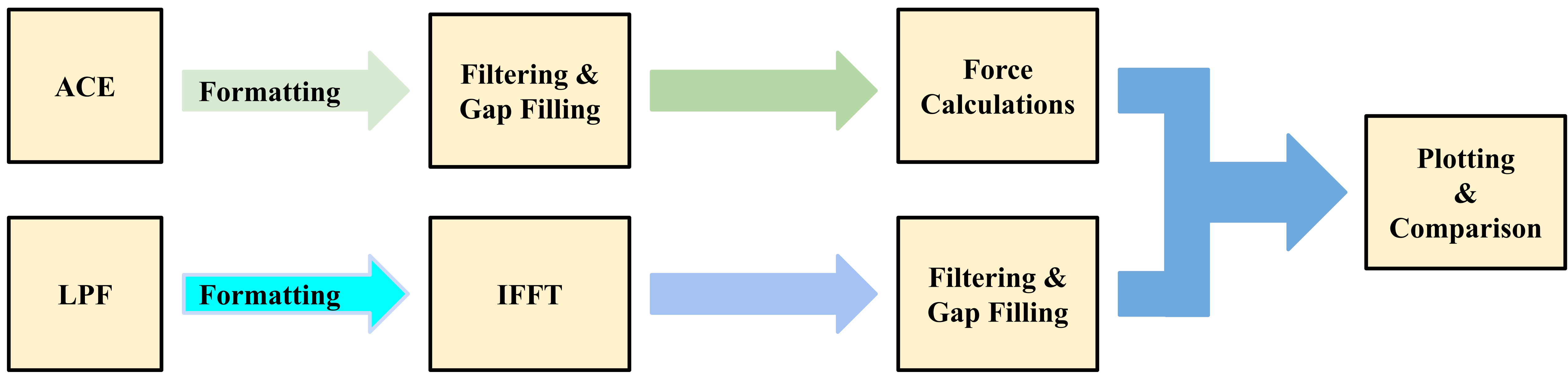}}
\caption{This figure shows the steps undertaken to reach two comparable data sets. Both datasets are converted into a readable format, and ACE's time frame is cut down to match that of LPF. LPF data was received in the frequency domain, so an inverse FFT was performed before filling in missing data blocks. ACE data is in time-series format, bad data (as indicated by code specified in the data header) is replaced and then the ACE data is modeled onto the LPF through force calculations.}

\label{Figure 2}
\end{figure}

\subsection{Process for Creating Plottable ACE Data}
Solar-wind data was obtained from the Advanced Composition Explorer (ACE). ACE data included information on solar-wind velocity components $(x, y, z)$, alpha particle to proton ratios, proton densities, and proton speeds at 64-second intervals measured in Coordinated Universal Time (UTC) \cite{aceswepamlevel2data}. The area of the solar array was modeled as a 2.9 meter diameter flat circular array, based off of the concept S/C design \cite{amaro2017laser}. ACE's UTC measurements were converted to GPS time because the LPF data is in GPS time. Bad data values from the ACE data were also removed (the ACE data header specifies that bad data is indicated by the value -9999.9) \cite{aceswepamlevel2data}. The gaps created by removing the bad data were split into three categories based on the process described in \textit{Modeling Spurious Forces on the LISA Spacecraft Across a Full Solar Cycle }\cite{frank2020modeling}. Type A gaps are missing one data point and are filled using linear interpolation. Type B gaps are missing two to twenty-four data points and are filled using linear interpolation with Gaussian Noise. Eq. 1 derives Gaussian Noise (\(\sigma\)) where \(\sigma_{1}\) and \(\sigma_{2}\) are the standard deviations found by sampling data of equal length to the gap on either side of the ACE data gap. In the equation, for the \(i\)th entry in a gap of length \(l\) such that \(1 \leq i \leq l\), a point was generated based on \(\sigma_{1}\) and \(\sigma_{2}\) \cite{frank2020modeling}. \\

\begin{equation}\label{eq:Gaussian Window Equations}
\sigma=\frac{l-(i-1)}{l+1}\sigma_{1}+\frac{i}{l+1}\sigma_{2} \label{eq1}
\end{equation}

Type C gaps are missing twenty-five or more data points; they are filled using a Hann window with a slow fall-off to zero, as defined by Eq. 2 \cite{frank2020modeling}. Twenty-five data points on either side of the gap are used to calculate simulated interpolation data \(\omega(n)\) where \textit{N}=51 and 0$<$\textit{n}$<$\textit{N}-1, creating the data to fill a type C gap.\\
\begin{equation}\label{eq:Hann Window}
\omega(n)=0.5\left[1-\cos{\left(\frac{2\pi n}{N-1}\right)}\right] \label{eq2}
\end{equation}

\begin{table}[htbp]
\caption{Gap-fill methods by gap length.
This set of methods is adapted from a previous publication. \cite{frank2020modeling}}
\begin{center}
\renewcommand{\arraystretch}{1.5}
\begin{tabular}{|c|c|c|}
\hline
Type & Length of Gap & Fill Method\\
\hline
A & 1 & Linear Interpolation\\
\hline
B & 2-24 & Gaussian Noise with Linear Interpolation\\
\hline
C & $\geq$25 & Hann Window\\
\hline
\end{tabular}
\label{tab1}
\end{center}
\end{table}

Following previous methods \cite{frank2020modeling}, the force exerted on the satellite by solar-wind particles was calculated using ACE data.\\
Eq. 3 calculates the number of protons (\(N_{p}\)) and alpha particles (\(N_{\alpha}\)) hitting the satellite per unit of time,\\
\begin{equation}\label{eq:Number of Particles Equation}
N=n v A\cos(\phi) \label{eq3}
\end{equation}
with $n$ being particle density and $v$ being wind speed. The surface area of the LPF solar array is represented by A. R is the fraction of particles reflected. \(\phi\) represents the angle between the normal of the array and the orbital plane \cite{frank2020modeling}. LPF's eta sensor data of the second test mass showed \(\phi\) was less than \(\pm\)$1 \times 10^{-7}$ for all values, making it reasonable to assume \(\phi\) in that case to be 0. The eta sensor data was stated to be within a 2-degree margin of error. Therefore, both a 0-degree and 2-degree value for the angle were tested. Equations 4,5,6 were used to calculate the final forces for the x and z axis. The $y$-axis equation was disregarded as the amount of force occurring on that axis was negligible. \cite{frank2020modeling}.

The following equations are reproduced verbatim from previous LISA analysis \cite{frank2020modeling}.
\begin{eqnarray}\label{eq:Force Equations}
F_{x} & = (N_{p}m_{p}+N_{\alpha}m_{\alpha})[(1+Rcos(2\phi))v_{x}+Rsin(2\phi)v_{z}], \\
F_{y} & = (N_{p}m_{p}+N_{\alpha}m_{\alpha})[(1-R)v_{y}], \\
F_{z} & = (N_{p}m_{p}+N_{\alpha}m_{\alpha})[(1+Rcos(2\phi))v_{z}+Rsin(2\phi)v_{x}] \label{eq456}
\end{eqnarray}

\(N_{p}\), the proton collision rate, and \(N_{\alpha}\), the alpha particle collision rate, were found using Eq. 3. The proton and alpha-particle masses from the ACE solar-wind data are represented by \(m_{p}\) and \(m_{\alpha}\). The $x, y, z$ components of solar-wind velocity are represented by \(v_{x},v_{y},\) and \(v_{z}\). Assuming the worst-case scenario (that all the particles reflect), R will be defined as 1. Then Equations 4,5,6 can be simplified to the following \cite{frank2020modeling}:
\begin{eqnarray}\label{eq:Simplified Force Equations}
F_{x} & = (N_{p}m_{p}+N_{\alpha}m_{\alpha})[(1+cos(2\phi))v_{x}+sin(2\phi)v_{z}], \\
F_{y} & = 0, \\
F_{z} & = (N_{p}m_{p}+N_{\alpha}m_{\alpha})[(1+cos(2\phi))v_{z}+sin(2\phi)v_{x}].\label{eq789}
\end{eqnarray}

The ACE solar-wind force values derived using Equations 7,8,9 are now able to be compared to the processed LPF data by being plotted and transformed as described in Section~III.

\subsection{Process for Creating Plottable LPF Data}
Data from the LPF was used to predict the effect of solar-wind on the future LISA data. The LPF data obtained from the NASA Goddard Space Flight Center combines information from various sensors to estimate the free-body motion of LPF. The data spans 114 days and is organized into 370 files, each 16384 seconds long. The data was taken at 2.5-second intervals from the second sensor on the $x$-axis and the second sensor on the $z$-axis. The data for LPF was in the frequency domain; thus, an IFFT was applied. The IFFT added the complex conjugates of negative values and zeros to the positive, non-zero values, returning the data to the original time series format. To calculate force, the acceleration from the IFFT was multiplied by the mass 422 kg \cite{2015micrometeroids}. The 31 gaps in both the $x$ and $z$ axes were filled using the Hann Window (Type C) method (Table. 1) \cite{frank2020modeling}.

\subsection{Plotting Methods}
A simple time domain comparison and a Fourier Transform comparison were used in Figures 3 and 4, respectively, to see if there were any apparent overlaps when the data were graphed over each other. The time domain data was zero-meaned to highlight the differences in the force data collected between LPF and ACE. The \texttt{scipy.signal.coherence()} function is used to determine coherence between LPF and ACE data and the averages of different lengths of correlation values \cite{2020SciPy-NMeth}. As described in the legends of Figures 5, 6, 7, and 8, the coherence comparison used increasing segment lengths multiplied by factors of 4 from 1 to \(4^{5}\).

\section{Results}
\begin{figure}[!ht]
\centerline{\includegraphics[width=\textwidth]{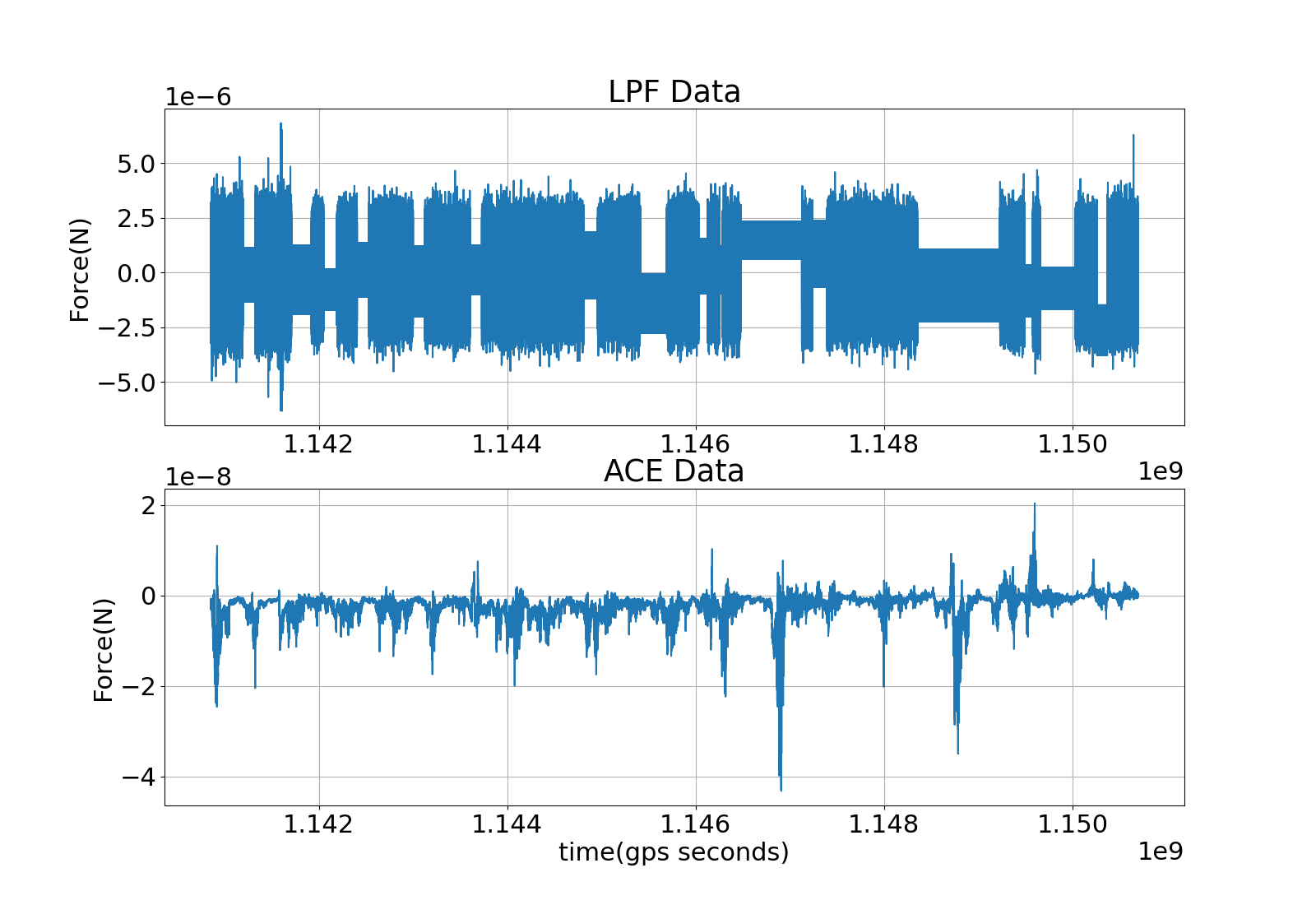}}
\caption{Time Domain plot of data for the $z$-axis. Data was zero-meaned and a 2-degree estimate was used for the ACE's force model.}
\label{Figure 3}
\end{figure}
Figure 3 depicts a time series that presents a direct comparison between the LPF time domain and the ACE force time domain, with both datasets having been zero-meaned. The ACE force data was calculated using the force equations and a 2-degree estimate for angle \(\phi\). The time series do not show any direct correlation between ACE and LISA data. Long constant segments in LPF indicate data gaps. \\

\begin{figure}[!ht]
\centerline{\includegraphics[width=\textwidth]{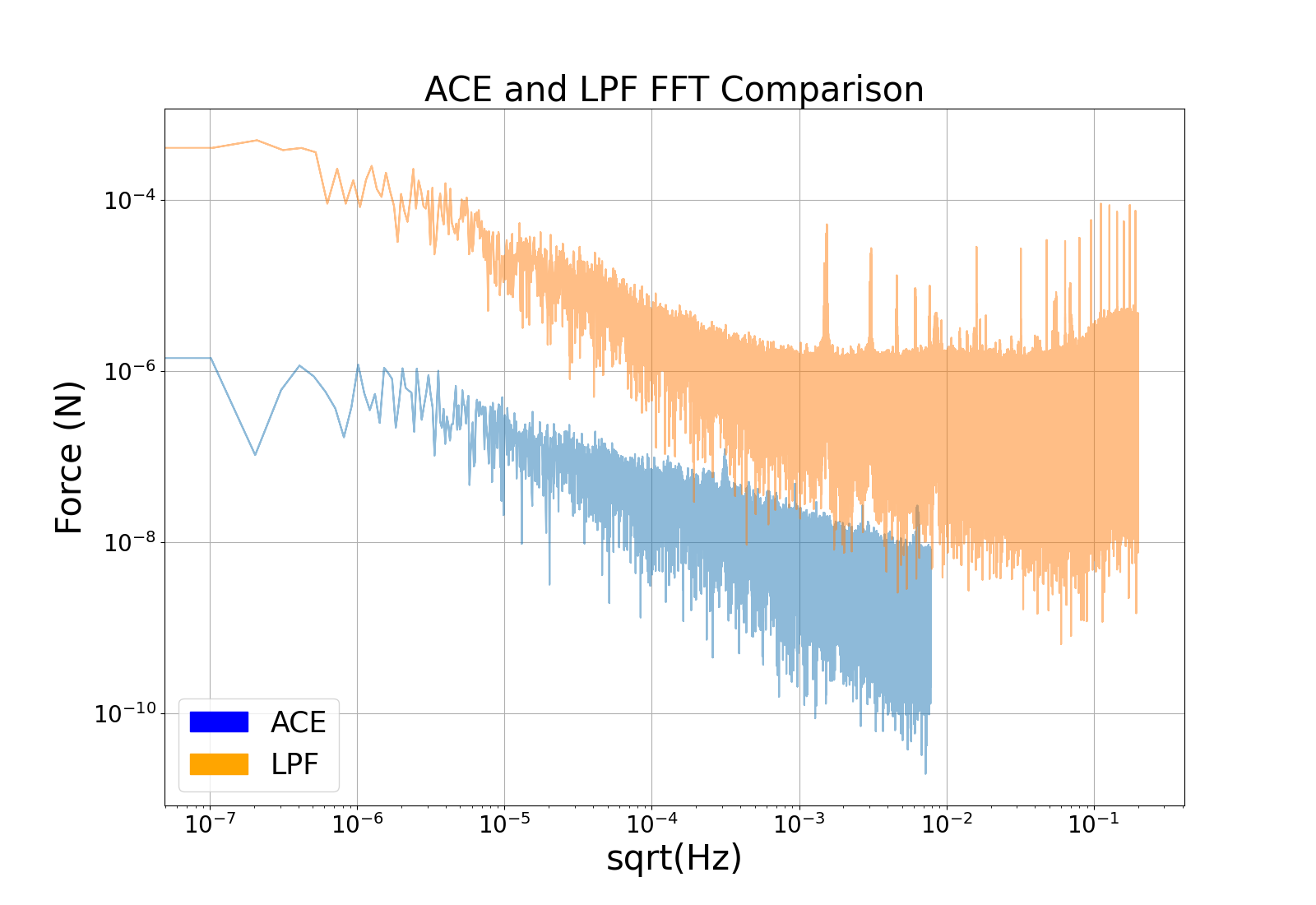}}
\caption{This is a Fourier-transform of the ACE and LPF time series from Figure 3. Despite the presence of spectral lines in both ACE and LPF, on closer inspection no coherent features are shown between the two spacecraft.}
\label{Figure 4}
\end{figure}
Figure 4 illustrates some overlap between the ACE and LPF data but nothing significant enough to be seen as a solid correlation between the two data sets. FFT was done with a 1/N normalization coefficient and a sampling time interval of dt = 2.5 seconds and dt = 64 seconds for LPF and ACE respectively.\\
\begin{figure}[!ht]
\centerline{\includegraphics[width=\textwidth]{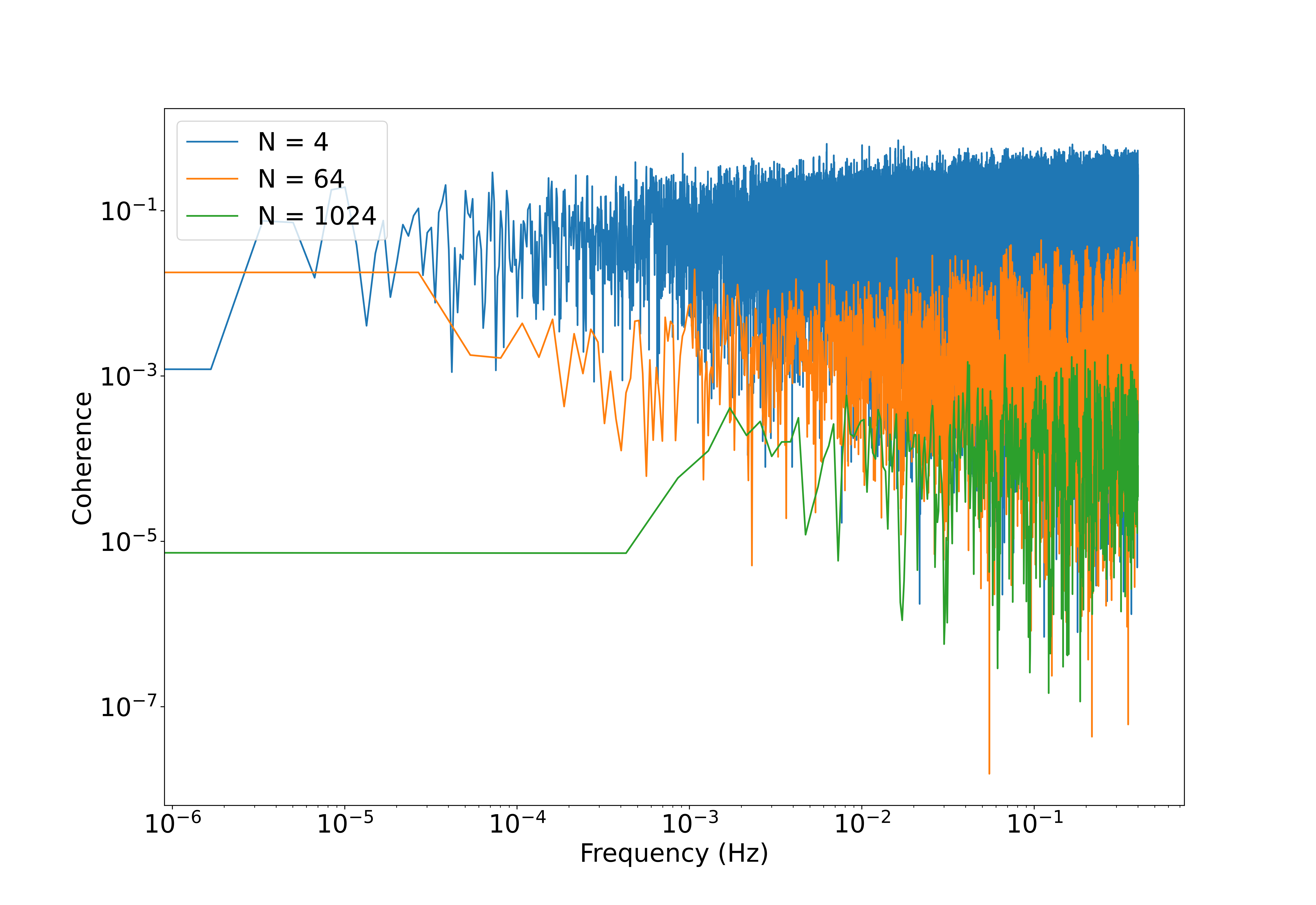}}
\caption{Coherence (frequency-domain correlation) between ACE and LPF data sets through six different lengths of coherence value averages for the $z$-axis. A 0-degree estimate was used for the ACE Force model. N is the number of coherence averages used. Variable coherence length was chosen for two reasons. First, low N was used to detect any possible low-frequency correlation. Second, increased N shows decreasing coherence, constraining any possible high-frequency correlation. Low levels of low-frequency correlation, below about 1 in 10 at 10 microHertz or 1 in 10000 at 10 millihertz, cannot be conclusively ruled out without longer-duration data.}
\label{Figure 5}
\end{figure}
From the coherence plot (Figure 5), it was observed that the coherence between the two data sets was minimal, and any correlation is probably due to random chance. In particular, spectral lines in LPF, due to instrumental artifacts, the line above 10 millihertz in ACE, do not show coherence on closer inspection.\\
\begin{figure}[!ht]
\centerline{\includegraphics[width=\textwidth]{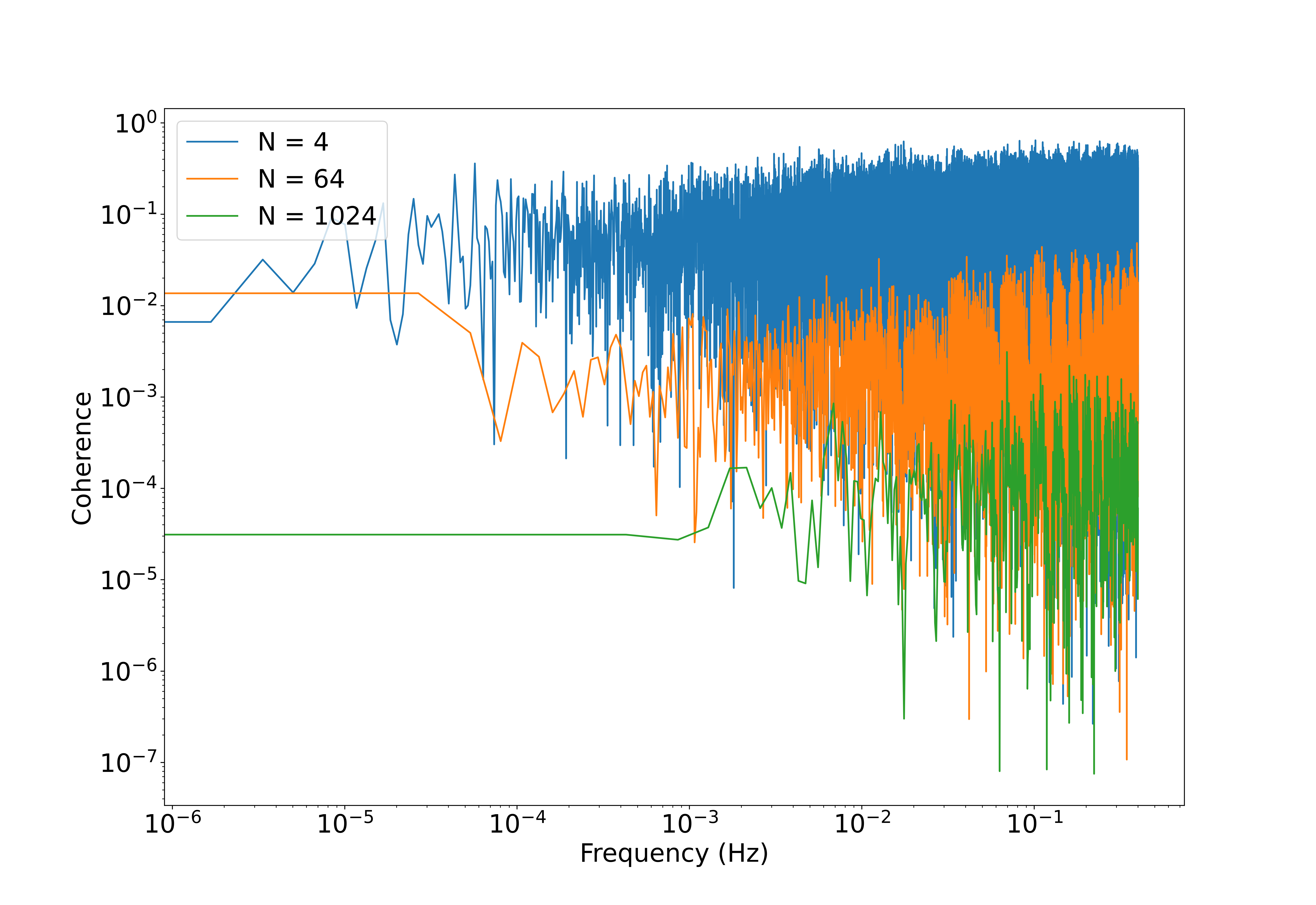}}
\caption{Coherence (frequency-domain correlation) between ACE and LPF data sets through six different lengths of coherence value averages for the $x$-axis. A 2-degree estimate was used for the ACE Force model. N is the number of coherence averages used.}
\label{Figure 6}
\end{figure}
Figure 5 shows six different lengths of coherence value averages. Each decrease in the values averaged increases the accuracy of the line. As the accuracy increases, the coherence decreases. The low coherence indicates that solar-wind will have little to no direct effects on the future LISA. Although with an estimated normal angle of 0, there should be no effect on the x-axis, the 2-degree margin of error of the angle was tested for coherence in the x-axis in Figure 6.\\

\section{Test Verification}
To ensure the accuracy of the previous tests and to create a visual of what would happen if there was a finding, the correlation methods were tested with simulated data. A measured known excitation of the sinusoidal form was added to both the LPF and ACE data sets after they had been filtered, gap-filled, and post-processed.

\begin{equation}\label{eq:Data Simulation}
F(t) = A \sin(2\pi f t+\phi) \label{eq6}
\end{equation}

In Eq. 10, F(t) represents the Force excitation added to the data where A is amplitude, f is frequency, t is time, and \(\phi\) is phase. The phase was set to 0 because it does not affect the coherence measurement. Amplitude and frequency were varied as follows. Adding this excitation in both sets of data simulates a specific frequency of the wave that should then be picked up by the coherence function, thus showing an excitation in that frequency on the coherence plot.

In each test, a single sinusoidal frequency was excited. Multiple frequencies were excited in independent tests. The test frequencies $1 \times 10^{-2}$ Hz, $1 \times 10^{-3}$ Hz, $2 \times 10^{-3}$ Hz, $5 \times 10^{-3}$ Hz, $1 \times 10^{-4}$ Hz, $5 \times 10^{-4}$ Hz were measured. All frequencies displayed distinct excitation at the injected frequency in the coherence spectrum. They were also tested for the least excitation needed for at least 50\% correlation when 1024 points were used per segment.
To illustrate this, Figure 7 shows the frequency $2 \times 10^{-3}$ Hz at 99.5\% coherence for 1024 segments, and Figure 8 shows the frequency $5 \times 10^{-3}$ Hz at 68.36\% coherence for 1024 segments per average with only 0.2 nanoNewtons of force. Figure 9 shows the frequency $5 \times 10^{-3}$ Hz again, but with the same 5 nanoNewtons of force as Figure 7 and displays a similar 99.9\% correlation. Thus showing that the coherence function works accurately and can properly compare the 2 datasets.\\

\begin{figure}[!ht]
\centerline{\includegraphics[width=\textwidth]{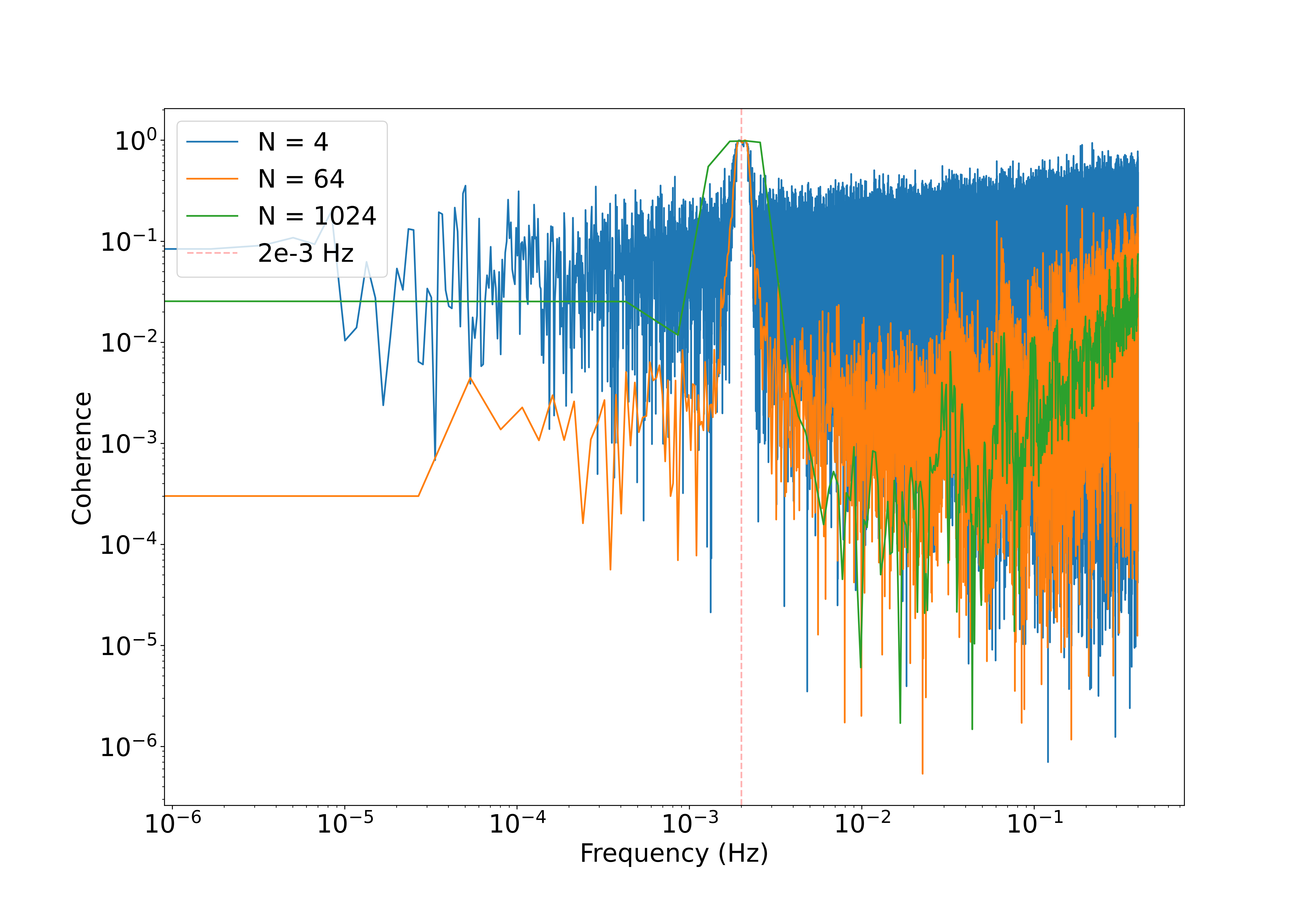}}
\caption{Coherence test of data with simulated frequency at $2 \times 10^{-3}$ Hz and amplitude of 5 nanoNewtons. N is the number of segments for coherence used. At 1024 points per segment, there is ~99.54\% correlation. }
\label{Figure 7}
\end{figure}
\begin{figure}[!ht]
\centerline{\includegraphics[width=\textwidth]{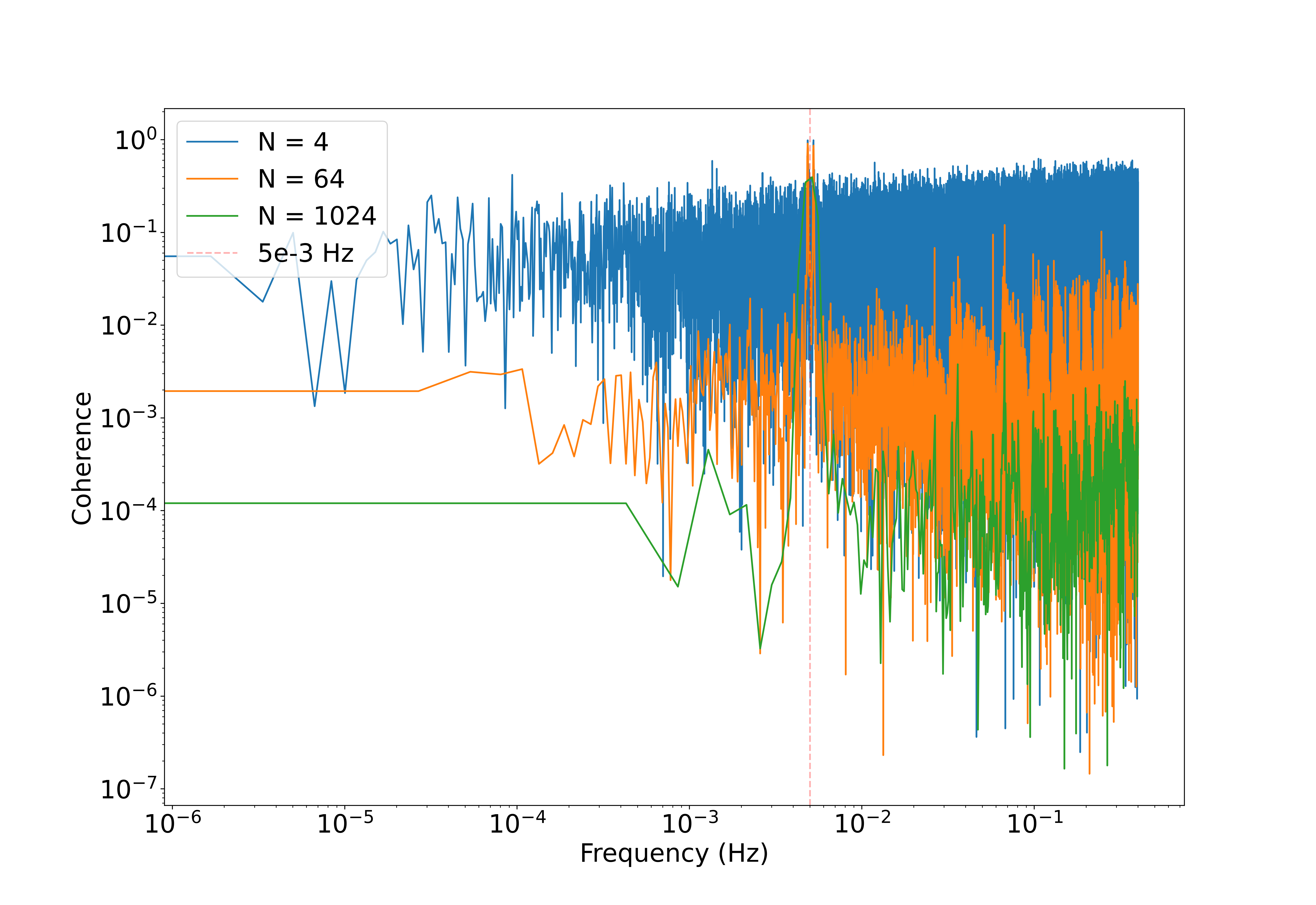}}
\caption{Coherence test of data with simulated frequency at $5 \times 10^{-3}$ Hz and amplitude of 0.2 nanoNewtons. N is the number of segments for coherence used. At 1024 points per coherence segment, there is ~68.36\% correlation. }
\label{Figure 8}
\end{figure}

\begin{figure}[!ht]
\centerline{\includegraphics[width=\textwidth]{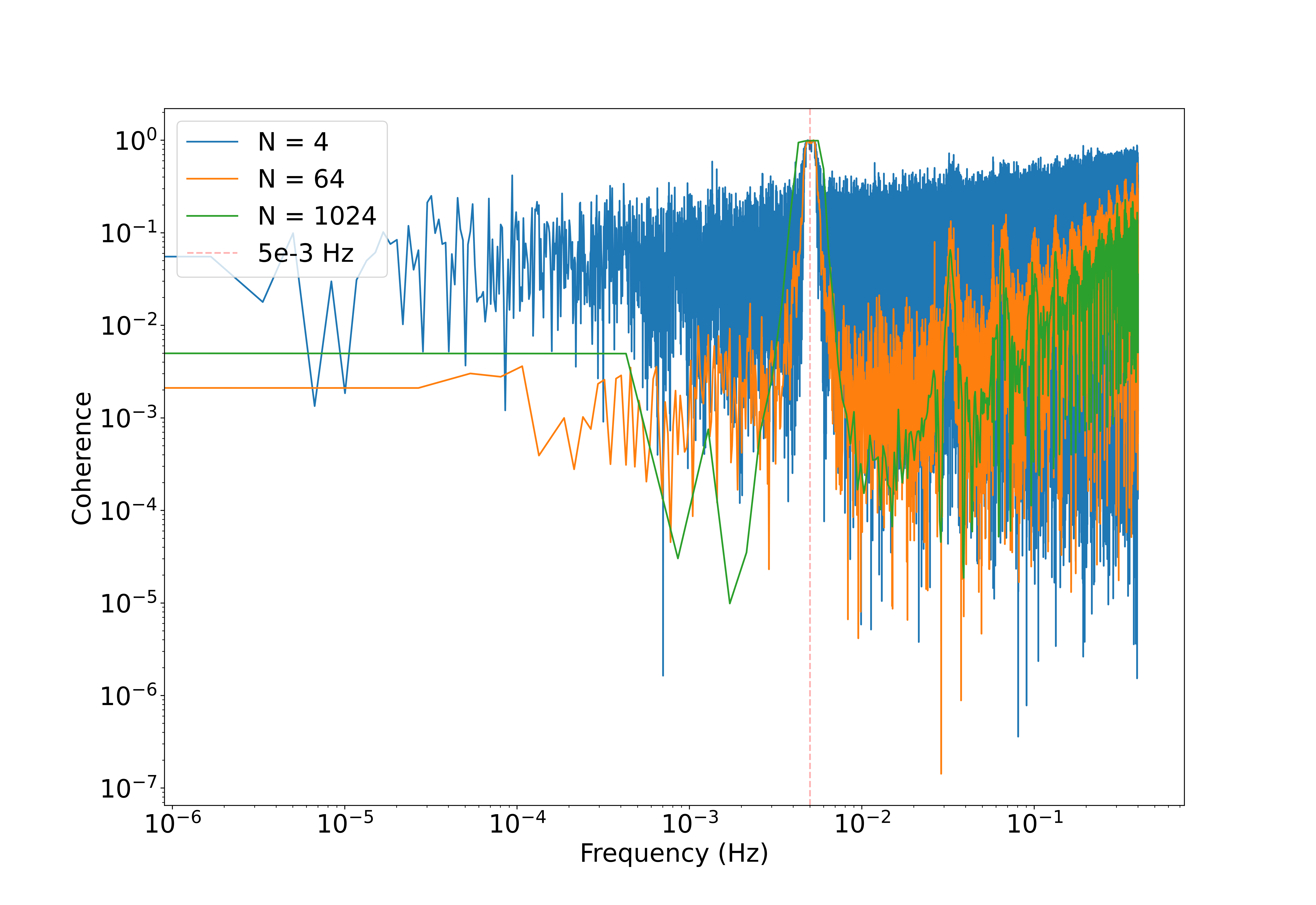}}
\caption{Coherence test of data with simulated frequency at $5 \times 10^{-3}$ Hz and amplitude of 5 nanoNewtons. N is the number of segments for coherence used. At 1024 points per coherence segment, there is ~99.9\% correlation. }
\label{Figure 9}
\end{figure}

The excitation frequencies of 2 and 5 millihertz in this paper represent the low-to-mid range of LISA, relevant to both the galactic white-dwarf binary population and black-hole mergers around a million solar masses. At these frequencies, prior work \cite{frank2020modeling} has estimated the force coupling of solar wind pressure into LISA: Figure 3 of that paper shows forces in the year 2000, near solar maximum, resulting in a predicted force at 2 millihertz of roughly 20 picoNewtons, and less at 5 millHertz. While a comprehensive statistical sampling from that work’s force spectrum is reserved for future research, this paper shows that near 200 picoNewtons, magnitude-squared coherence levels do reach above 0.68, or about two-thirds, at 5 millihertz. Comparing the two frequencies, the coherence would be expected to be lower at 2 millihertz, consistent with the noise spectrum.

\section{Conclusion}
The effects of spurious solar-wind on acceleration noise of LISA Pathfinder (LPF) data from the $x$ and $z$ axes were thoroughly analyzed. As shown in Figures 5 and 6, the coherence between the Advanced Composition Explorer (ACE) and LPF data was sporadic for the given frequency range, and the average correlation for each data section was minuscule. The coherence test has verified that there is a measurable correlation for a simulated excitation of known amplitude, as seen in Figures 7, 8 and 9. The analysis shows that the correlation seen in the real, unexcited data is indistinguishable from chance given the data available and no relationship was found between ACE solar-wind data and LPF data. The models and plots of the $x$ and $z$ axes found a negligible relation between the ACE solar-wind data set and the LPF.

This result suggests that there is a reasonable safety margin between solar wind forces and measurable acceleration noise in LISA. More detailed analysis is warranted, as solar variability could reach levels even higher than in 2000, but comparison with solar irradiance \cite{frank2020modeling} suggests that those forces may pose a larger concern. To the benefit of LISA, albeit a loss to the space weather community, it currently appears difficult, with such low coherence at expected force levels, to solve the inverse problem of finding solar wind force from LISA acceleration, as proposed previously \cite{shaul2006solar}.

These findings are encouraging for the future of the LISA as less noise will allow LISA to see much more of the universe. However, solar irradiance data, another noise factor with likely greater effects on the LPF, has yet to be tested \cite{frank2020modeling}, and the limited 3-month time frame of data that were analyzed means that correlations over longer timescales are possible as well. Future analyses should also focus on working with individual subsets of the data and plotting the relationships between those to eliminate the possible confounding effects created when using gap-fill methods. \\

\section{Acknowledgments}
Thank you to the Institute for Computing in Research: Mark Galassi and Rhonda Crespo. Thank you I. Thorpe and J. Slutsky at the NASA Goddard Space Flight Center for the data and helpful comments. Additionally, thank you to B.M. Frank, B. Piotrzkowsky, B. Bolen, M. Cavaglià, and S.L. Larson for their methods and helpful comments. Indie Desiderio-Sloane and Arnold Yang were supported by the Institute for Computing in Research. GDM was supported by multiple Los Alamos National Laboratory internal funding sources, including the Intelligence and Space Research Early Career (ISR-EC) 2022 Pitch Day award XB4F00 and the Center for Space and Earth Science project development award W88600. This document has Los Alamos Unlimited Release (LA-UR) number LA-UR-23-30801. This work was supported by the U.S. Department of Energy through the Los Alamos National Laboratory. Los Alamos National Laboratory is operated by Triad National Security, LLC, for the National Nuclear Security Administration of U.S. Department of Energy (Contract No. 89233218CNA000001).

\section{Supplementary Data}
The source code and procedures to run it are saved at codeberg.org/Poarthan/solarWind- DataAnalysisCode. The ACE and LPF data used can be found respectively at zenodo.org/records/6955182 and zenodo.org/records/6954044.

\section*{References}
\bibliographystyle{iopart-num}
\bibliography{solarWind.bib}

\providecommand{\noopsort}[1]{}\providecommand{\singleletter}[1]{#1}%
\providecommand{\newblock}{}
\begin{thebibliography}{10}
\expandafter\ifx\csname url\endcsname\relax
  \def\url#1{{\tt #1}}\fi
\expandafter\ifx\csname urlprefix\endcsname\relax\def\urlprefix{URL }\fi
\providecommand{\eprint}[2][]{\url{#2}}

\bibitem{amaro2017laser}
Amaro-Seoane P, Audley H, Babak S, Baker J, Barausse E, Bender P, Berti E,
  Binetruy P, Born M, Bortoluzzi D {\em et~al.\/} 2017 {\em arXiv preprint
  arXiv:1702.00786\/}

\bibitem{LIGO2015}
Collaboration T~L~S, Aasi J, Abbott B~P, Abbott R, Abbott T, Abernathy M~R,
  Ackley K, Adams C, Adams T, Addesso P and et~al 2015 {\em Classical and
  Quantum Gravity\/} {\bf 32} 074001

\bibitem{shaul2006solar}
Shaul D, Aplin K, Araujo H, Bingham R, Blake J, Branduardi-Raymont G, Buchman
  S, Fazakerley A, Finn L, Fletcher L {\em et~al.\/} 2006 Solar and cosmic ray
  physics and the space environment: Studies for and with {LISA} {\em AIP
  Conference Proceedings\/} vol 873 (American Institute of Physics) pp 172--178
  \urlprefix\url{https://doi.org/10.1063/1.2405038}

\bibitem{frank2020modeling}
Frank B~M, Piotrzkowski B, Bolen B, Cavagli{\`a} M and Larson S~L 2020 {\em
  Classical and Quantum Gravity\/} {\bf 37} 175007

\bibitem{kaieser-et-al-2008}
Guhathakurta M~L~K~~T~A~K~~J~M~D~~O~C~S~C~~M and Christian E 2008 {\em Space
  Science Reviews\/}
  \urlprefix\url{https://ui.adsabs.harvard.edu/abs/2008SSRv..136....5K/abstract}

\bibitem{arge1999}
Arge C~N and Pizzo V~J 2000 {\em Journal of Geophysical Research: Space
  Physics\/} {\bf 105} 10465--10479 (\textit{Preprint}
  \eprint{https://agupubs.onlinelibrary.wiley.com/doi/pdf/10.1029/1999JA000262})
  \urlprefix\url{https://agupubs.onlinelibrary.wiley.com/doi/abs/10.1029/1999JA000262}

\bibitem{aceswepamlevel2data}
 2016 {ACE} {SWEPAM} level 2 data
  \urlprefix\url{https://izw1.caltech.edu/ACE/ASC/level2/lvl2DATA\_SWEPAM.html}

\bibitem{thorpe2019micrometeoroid}
Thorpe J~I, Slutsky J, Baker J~G, Littenberg T~B, Hourihane S, Pagane N,
  Pokorny P, Janches D, Armano M, Audley H {\em et~al.\/} 2019 {\em The
  Astrophysical Journal\/} {\bf 883} 53

\bibitem{Grimani_2006}
Grimani C, Bagni G, Fabi M, Vicerè A, Marconi L, Stanga R, Bosi L, Vocca H,
  Araújo H, Shaul D, Sumner T, Wass P, Boatella C, Lobo A, Chmeissani M and
  Martinez I 2006 {\em Journal of Physics: Conference Series\/} {\bf 32} 6
  \urlprefix\url{https://dx.doi.org/10.1088/1742-6596/32/1/002}

\bibitem{10.1093/mnras/stac316}
Kumar P, White S~M, Stovall K, Dowell J and Taylor G~B 2022 {\em Monthly
  Notices of the Royal Astronomical Society\/} {\bf 511} 3937--3950 ISSN
  0035-8711 (\textit{Preprint}
  \eprint{https://academic.oup.com/mnras/article-pdf/511/3/3937/42579069/stac316.pdf})
  \urlprefix\url{https://doi.org/10.1093/mnras/stac316}

\bibitem{smetana2020background}
Smetana A 2020 {\em Monthly Notices of the Royal Astronomical Society:
  Letters\/} {\bf 499} L77--L81

\bibitem{harris2020array}
Harris C~R, Millman K~J, van~der Walt S~J, Gommers R, Virtanen P, Cournapeau D,
  Wieser E, Taylor J, Berg S, Smith N~J, Kern R, Picus M, Hoyer S, van Kerkwijk
  M~H, Brett M, Haldane A, del R{\'{i}}o J~F, Wiebe M, Peterson P,
  G{\'{e}}rard-Marchant P, Sheppard K, Reddy T, Weckesser W, Abbasi H, Gohlke C
  and Oliphant T~E 2020 {\em Nature\/} {\bf 585} 357--362
  \urlprefix\url{https://doi.org/10.1038/s41586-020-2649-2}

\bibitem{2015micrometeroids}
Thorpe J, Parvini C and Trigo-Rodríguez J 2015 {\em Astronomy \&
  Astrophysics\/} {\bf 586}

\bibitem{2020SciPy-NMeth}
Virtanen P, Gommers R, Oliphant T~E, Haberland M, Reddy T, Cournapeau D,
  Burovski E, Peterson P, Weckesser W, Bright J, {van der Walt} S~J, Brett M,
  Wilson J, Millman K~J, Mayorov N, Nelson A~R~J, Jones E, Kern R, Larson E,
  Carey C~J, Polat {\.I}, Feng Y, Moore E~W, {VanderPlas} J, Laxalde D,
  Perktold J, Cimrman R, Henriksen I, Quintero E~A, Harris C~R, Archibald A~M,
  Ribeiro A~H, Pedregosa F, {van Mulbregt} P and {SciPy 10 Contributors} 2020
  {\em Nature Methods\/} {\bf 17} 261--272

\end{thebibliography}

\end{document}